\documentclass[aps,prl,twocolumn,floatfix,showpacs,preprintnumbers,amsmath,amssymb]{revtex4-1}
\usepackage[dvips]{graphics}

\usepackage{psfrag}
\usepackage{graphicx}
\usepackage{trfsigns}
\usepackage{amssymb}
\usepackage[usenames,dvipsnames]{color}
\usepackage{dashrule}
\usepackage{textgreek}
\usepackage[english]{babel}
\usepackage{contour}
\usepackage{upgreek,bm}
\usepackage{color}
\usepackage{dsfont}
\usepackage{soul}
\usepackage{braket}
\usepackage{cancel}
\usepackage{subfigure}
\usepackage{overpic}

\definecolor{dred}{rgb}{.6,.0,0.}
\definecolor{dblue}{rgb}{.0,.0,0.6}

\renewcommand{\vec}[1]{\mathbf{#1}}

\newcommand{\tens}[1]{\mbox{\textsf{\textbf{#1}}}}

\newcommand{\sprod}{\!\cdot\!}

\newcommand{\dif}{\mathrm{d}}
\newcommand{\mi}{\textrm{i}} 
\newcommand{\me}{\mathrm{e}}

\usepackage{todonotes}
\presetkeys{todonotes}{fancyline, color=white}{}

\begin{document}

\title{Non-additivity of optical and Casimir--Polder potentials}

\author{Sebastian Fuchs$^1$}
\author{Robert Bennett$^1$}
\author{Roman V. Krems$^2$}
\author{Stefan Yoshi Buhmann$^{1,3}$}
\affiliation{$^1$ Physikalisches Institut, Albert-Ludwigs-Universit\"at Freiburg, Hermann-Herder-Stra{\ss}e 3, 79104 Freiburg, Germany\\
$^2$ Department of Chemistry, University of British Columbia, Vancouver, British Columbia, V6T 1Z1, Canada\\
$^3$ Freiburg Institute for Advanced Studies, Albert-Ludwigs-Universit\"at Freiburg, Albertstra{\ss}e 19, 79104 Freiburg, Germany}

\date{\today}

\begin{abstract}
An atom irradiated by an off-resonant laser field near a surface is expected to experience the sum of two fundamental potentials, the optical potential of the laser field and the Casimir--Polder potential of the surface. Here, we report a new non-additive potential, namely the laser-induced Casimir--Polder potential, which arises from a correlated coupling of the atom with both the laser and the quantum vacuum. We apply this result to an experimentally realizable scenario of an atomic mirror with an evanescent laser beam leaking out of a surface. We show that the non-additive term is significant for  realistic experimental parameters, transforming potential barriers into potential wells, which can be used to trap atoms near surfaces. 
\end{abstract}

\maketitle
In experiments involving an applied electromagnetic field and particles trapped in or near material objects, two forces of very different origin arise. One is the Casimir--Polder (CP) force \cite{Casimir_Polder:1948} arising from the interaction between an atom and the electromagnetic vacuum field, which is restricted and modified by the presence of nearby surfaces. The other is the optical force stemming from the direct interaction between the applied electromagnetic field and the atom. The latter was first applied in experiments where micron-size dielectric spheres were trapped by two laser beams \cite{Ashkin:1970}, eventually leading to the first observation of optical trapping of atoms by a single strongly focused Gaussian laser beam \cite{Chu:1986}. Atoms can also be reflected by an evanescent laser field at a surface, as shown for example in Ref.~\cite{Balykin:1988} where state-selective reflection of Na atoms was demonstrated using an evanescent field.

The CP force has been studied extensively. It is well described by many established theoretical approaches, ranging from a quantum-mechanical linear-response formalism \cite{Wylie:1985, Wylie:1984} to macroscopic extensions of quantum electrodynamics that incorporates material properties, e.g. Ref.~\cite{Scheel:2008}. There are also several experimental techniques to investigate this force. Almost fifty years after its theoretical prediction, the CP force was first measured by observing the angle of deflection of atoms passing through a V-shaped cavity \cite{Sukenik:1993}. Later, it was demonstrated that the scattering of slowly-moving atoms from a surface can also be used to deduce the CP potential \cite{Shimizu:2001}. Recently, approaches involving a single Rb atom optically trapped close to a surface have been employed \cite{Bender:2010}. There, the sum of the trapping potential and the CP potential determines the equilibrium position of the atom, so that for a known trapping potential the CP potential can be determined.

In a recent experiment this technique has been used to make a first direct measurement of CP forces between solid surfaces and atomic gases in the transition regime between short distances (non-retarded) and long distances (retarded) \cite{Bender:2010, Bender:2014}. Here, ultracold ground-state Rb atoms are reflected from an evanescent wave barrier at a glass prism.

The question arises whether the two fundamentally different ingredients, the CP potential ($U_{\textrm{CP}}$) and the light-induced optical potential ($U_{\textrm{L}}$), can be simply added to obtain the total potential.  Being inherently related to universal scaling laws of dispersion potentials \cite{Buhmann:2010}, this subject is of fundamental relevance. In this paper we report a new nonadditive laser-induced CP potential. We show that it plays an important role under specific experimental conditions. 

The CP potential can be viewed as the modification of a fluctuating dipole moment due to the electric field those fluctuations induce. It is clear that this is a recursive process; the dipole induces a field, which in turn interacts with the dipole, changing the field it produces, which then interacts again with the dipole and so on. Normally, only the first step in this process is considered where the fluctuations of the dipole moment are taken to come from the field that would be present if the dipole were not there. This electric field could come from the vacuum field or a laser field depending on the specific context. This leads to the usual approach taken when considering atoms subject to vacuum and laser fields; the two potentials are simply added. Here we take the next step, in which the laser light effect on the dipole moment is taken into account when calculating the CP potential.

\begin{figure}
\centerline{\includegraphics[width=\columnwidth]{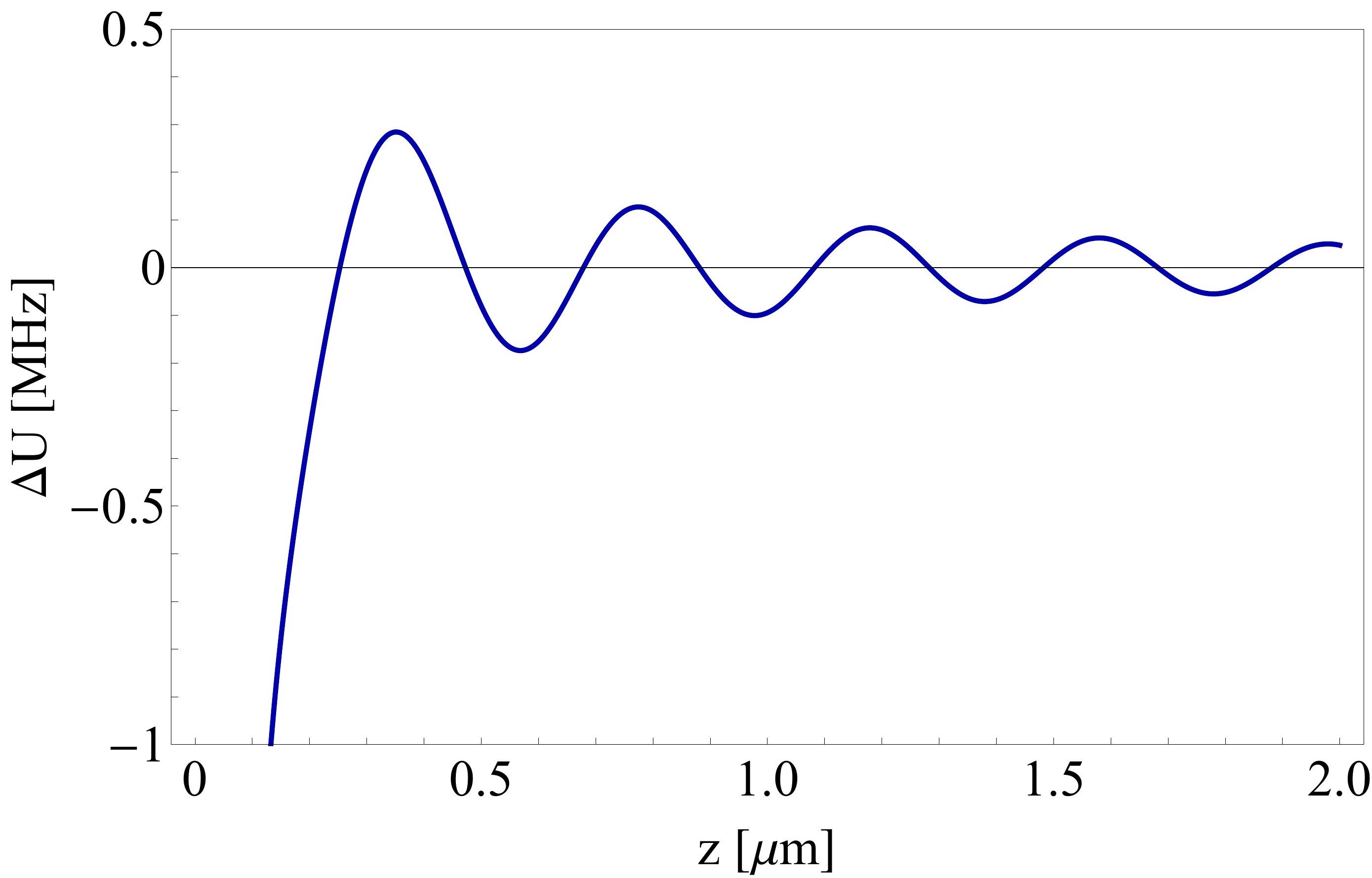}}
\caption{Difference $\Delta U = U_{\textrm{CP}} - U_{\textrm{LCP}}$ for a perfectly conducting mirror with reflective coefficients $r_{\textrm{s}} = -1$ and $r_{\textrm{p}} = 1$. The laser intensity is $I=5 \textrm{W/cm}^2$ used for Rb atoms with a transition frequency of $\tilde{\omega}_{10} = 2.37 \times 10^{15}$ rad/s. The detuning is $\Delta = 2 \pi \times 10^8$ rad/s. The angle between between the z-axis and the orientation of the field $\vec{E} \left( \vec{r}_{\textrm{A}} \right)$ is $\theta = \pi/2$.}
\label{fig:PlotTwo}
\end{figure}

Quantum vacuum forces between an atom in its ground state and a surface are attractive and in general not suitable to create a stable position for the atom \cite{Rahi:2010}. Nevertheless, Ref.~\cite{Chang:2013} describes the possibility of trapping ground-state atoms by dressing them with an excited state whose potential is repulsive in a laser field. This method is similar in spirit to our proposal, especially because the laser-induced CP potential for the atomic ground state strongly resembles the excited-state CP potential (cf. Fig.~\ref{fig:PlotTwo}).

We describe the system using a Hamiltonian that governs the coupling of the atom to the electric field and consequently consists of a field part, an atomic part and a multipolar dipole-field coupling. The field Hamiltonian $\hat{H}_{\textrm{F}}$ can be expressed as 
\begin{equation}
\hat{H}_{\textrm{F}} = \sum\limits_{\lambda = \textrm{e}, \textrm{m}} \int \dif^3 r \int\limits^{\infty}_0{\dif \omega \hbar \omega \, \hat{\vec{f}}^{\dagger}_{\lambda} \left( \vec{r}, \omega \right) \sprod \hat{\vec{f}}{}_{\lambda} \left( \vec{r}, \omega \right)},
\label{eq:Field Hamiltonian}
\end{equation}
where $\hat{\vec{f}}_{\lambda}$ and $\hat{\vec{f}}^{\dagger}_{\lambda}$ are creation and annihilation operators for composite field-matter excitations. Here we model a driving laser as being a result of a source occupying a volume $V_{\textrm{S}}$ in space. This leads us to represent field states as a product
\begin{equation}
\ket{\psi}_{\textrm{F}} = \underset{\vec{r} \in V_{\textrm{S}}}{\ket{\left\{ \vec{f}_{\lambda} \left( \vec{r}, \omega \right) \right\}}} \otimes \underset{\vec{r} \notin V_{\textrm{S}}}{\ket{ \left\{ 0 \right\}}}
\label{eq:Quantum State}
\end{equation}
of coherent excitations $\ket{ \{ \vec{f}_{\lambda} \left( \vec{r}, \omega \right) \}}$ in the source region and the vacuum state $\ket{\left\{ 0 \right\}}$ for all other regions. If the annihilation operator $\hat{\vec{f}}_{\lambda}\left( \vec{r}, \omega \right)$ acts on the state \eqref{eq:Quantum State}, there are consequently two contributions
\begin{equation}
\hat{\vec{f}}_{\lambda} \left( \vec{r}, \omega \right) \ket{\psi}_{\textrm{F}} = \begin{cases} \vec{f}_{\lambda} \left( \vec{r}, \omega \right) \ket{\psi}_{\textrm{F}} & \textrm{if} \; \vec{r} \in V_{\textrm{S}}\\ 0 & \textrm{if} \; \vec{r} \notin V_{\textrm{S}} \end{cases}.
\label{eq:Electric Field Quantum State}
\end{equation}
The atom-field Hamiltonian $\hat{H}_{\textrm{AF}} = - \hat{\vec{d}} \sprod \hat{\vec{E}} \left( \vec{r}_{\textrm{A}} \right)$ in the multipolar coupling scheme is determined by the electric field at the atom's position $\vec{r}_{\textrm{A}}$ and the dipole operator $\hat{\vec{d}}$. The electric field is given by the respective classical Green's tensor $\tens{G} \left( \vec{r}, \vec{r}{}_{\textrm{A}}, \omega \right)$ and the field operator $\hat{\vec{f}}_{\lambda} \left( \vec{r}, \omega \right)$. Solving the Heisenberg equation of motion for the field operator using the field Hamiltonian $\hat{H}_{\textrm{F}}$ \eqref{eq:Field Hamiltonian} and the coupling Hamiltonian $\hat{H}_{\textrm{AF}}$ and inserting this back into the E field, the final expression for the time-dependent electric field operator yields
\begin{align}\label{eq:Electric Field Free and Induced}
\hat{\vec{E}} \left( \vec{r}, \omega, t \right) &= \hat{\vec{E}}{}_{\textrm{free}} \left( \vec{r}, \omega, t \right) + \hat{\vec{E}}{}_{\textrm{ind}} \left( \vec{r}, \omega \right)\notag\\
&= \hat{\vec{E}} \left( \vec{r}, \omega \right) \me^{- \mi \omega ( t - t_0 )}\notag \\
&\!\!\!\!\!\!\!+ \frac{\mi \mu_0}{\pi} \omega^2 \int\limits^t_{t_0}{\dif t' \me^{-\mi \omega ( t-t' )} \mathrm{Im} \tens{G} \left( \vec{r}, \vec{r}{}_{\textrm{A}}, \omega \right) \sprod \hat{\vec{d}} ( t' )}.
\end{align}
$\mu_0$ is the permeability of free space. The induced contribution represents the inhomogeneous part of the solution and couples the Green's tensor to the atomic dipole moment as shown in Ref.~\cite{Dalibard:1982}. The state \eqref{eq:Electric Field Quantum State} can be inserted into Eq.~\eqref{eq:Electric Field Free and Induced} where the free component is modeled as a classical laser driving field of frequency $\omega_{\textrm{L}}$ at the atom's position, $\vec{E} \left( \vec{r}_{\textrm{A}}, t \right) = \vec{E} \left( \vec{r}_{\textrm{A}} \right) \cos \left( \omega_{\textrm{L}} t \right)$.

In a similar way, one can compute the Heisenberg equation of motion for the atomic flip operator $\hat{A}{}_{mn} \left( t \right)$  \cite{Buhmann_Book_2} defined in such a way that the atomic part of the Hamiltonian is $\hat{H}_{\textrm{A}} = \sum_{n}{E_n \hat{A}_{nn}}$. The electric field \eqref{eq:Electric Field Free and Induced} is evaluated using the Markov approximation for weak atom-field coupling and we discard slow non-oscillatory dynamics of the flip operator by setting $\hat{A}{}_{mn} \left( t' \right) \simeq \me^{\mi \tilde{\omega}{}_{mn} \left( t'-t \right)} \hat{A}{}_{mn} \left( t \right)$ for the time interval $t_0 \leq t' \leq t$. To apply the Markov approximation we have assumed that the atomic transition frequency $\tilde{\omega}_{10}$ is not close to any narrow-band resonance mode of the medium. If there were such a mode, the atom would mostly interact with it, similar to a cavity. In this case the mode could be modeled by a Lorentzian profile \cite{Haroche:1991, Buhmann_Welsch:2008, Buhmann_Book_2}.

The parameters entering the dynamics are the shifted frequency $\tilde{\omega}{}_{mn} = \omega_{mn} + \delta \omega_{mn}$, where $\omega_{mn}$ is the atom's pure eigenfrequency and the CP frequency shift $\delta \omega_{mn}$ due to the presence of the surface is the rate of spontaneous emission $\Gamma_{mn}$. The fast-oscillating nondiagonal parts $\hat{A}{}_{mn} \left( t \right)$ can be decoupled from the slowly-oscillating diagonal operator terms $\hat{A}{}_{mm} \left( t \right)$ by assuming that the atom does not have quasi-degenerate transitions. Moreover the atom is unpolarized in each of its energy eigenstates so that $\vec{d}{}_{mm} = \vec{0}$, which is guaranteed by atomic selection rules \cite{Buhmann_Book_2}. Finally, we assume the atom stays in its initial state with $\langle \hat{A}_{kl} \left( t' \right) \rangle \approx \langle \hat{A}_{kl} \left( t \right) \rangle \approx \delta_{kn} \delta_{ln}$. Consequently, to compute the dipole moment we only need the nondiagonal elements of the atomic flip operator
\begin{multline}
\langle \dot{\hat{A}}_{mn} \left( t \right) \rangle = \mi \tilde{\omega}_{mn} \langle \hat{A}_{mn} \left( t \right) \rangle - \frac{1}{2} \left[ \Gamma_n + \Gamma_m \right] \langle \hat{A}_{mn} \left( t \right) \rangle\\
+\frac{\mi}{\hbar} \sum\limits_k{ \left[ \langle \hat{A}_{mk} \left( t \right) \rangle \vec{d}_{nk} - \langle \hat{A}_{kn} \left( t \right) \rangle \vec{d}_{km} \right] \cdot \vec{E} \left( \vec{r}_{\textrm{A}}, t \right)}.
\label{eq:Atomic Flip Operator Nondiagonal}
\end{multline}
Similar to the electric field \eqref{eq:Electric Field Free and Induced}, the first line of Eq.~\eqref{eq:Atomic Flip Operator Nondiagonal} for the atomic flip operator finds its way into the free part of the dipole moment $\hat{\vec{d}}_{\textrm{free}} \left( t \right)$, while the second part containing the electric field becomes the laser induced part $\hat{\vec{d}}_{\textrm{ind}} \left( t \right)$.

In the Markov approximation, we use the expression of the complex atomic polarizability for an atom in a spherically symmetric state with negligible damping
\begin{equation}
\alpha_n \left( \omega \right) = \frac{2}{3 \hbar} \sum\limits_k{\frac{\tilde{\omega}_{kn} \left| \vec{d}_{nk} \right|^2}{\tilde{\omega}^2_{kn} - \omega^2}} \tens{1},
\label{eq:Atomic Polarizability}
\end{equation}
where $\tens{1}$ is the unit matrix. The dipole moment in time domain reads
\begin{equation}
\langle \hat{\vec{d}}_{\textrm{ind}} \left( t \right) \rangle_n = \frac{1}{2} \left[ \alpha_n \left( \omega_{\textrm{L}} \right) \vec{E} \left( \vec{r}_{\textrm{A}} \right) \me^{-\mi \omega_{\textrm{L}}t} + \textrm{h.c.} \right]
\label{eq:Dipole Moment Time Domain}
\end{equation}
which oscillates with the laser frequency $\omega_{\textrm{L}}$. An equivalent expression for the induced electric field $\hat{\vec{E}}_{\textrm{ind}} \left( \vec{r}, t \right)$ is obtained in a similar way.

The quantity we are interested in is the total potential $U = - \frac{1}{2} \langle \hat{\vec{d}} \left( t \right) \sprod \hat{\vec{E}} \left( \vec{r}_{\textrm{A}}, t \right) \rangle$ which consists of various contributions as shown in Tab.~\ref{fig:Cascade}. At leading order, the induced part of the field (dipole) depends on the free part of the dipole moment (field), given by the first iteration of Eq.~\eqref{eq:Electric Field Free and Induced}. This leads to two well-known potentials, namely the laser-light potential and the standard, undriven CP potential. Going one iteration further yields an additional contribution, the CP potential under the influence of the driving laser field. This contribution is non-additive, i.e., the total potential experienced by the atom can no longer be obtained simply by summing the Casimir-Polder potential and the laser-induced potential. We will show that this non-additive contribution can be significant under certain circumstances.
\begin{table}
\centerline{\includegraphics[width=\columnwidth]{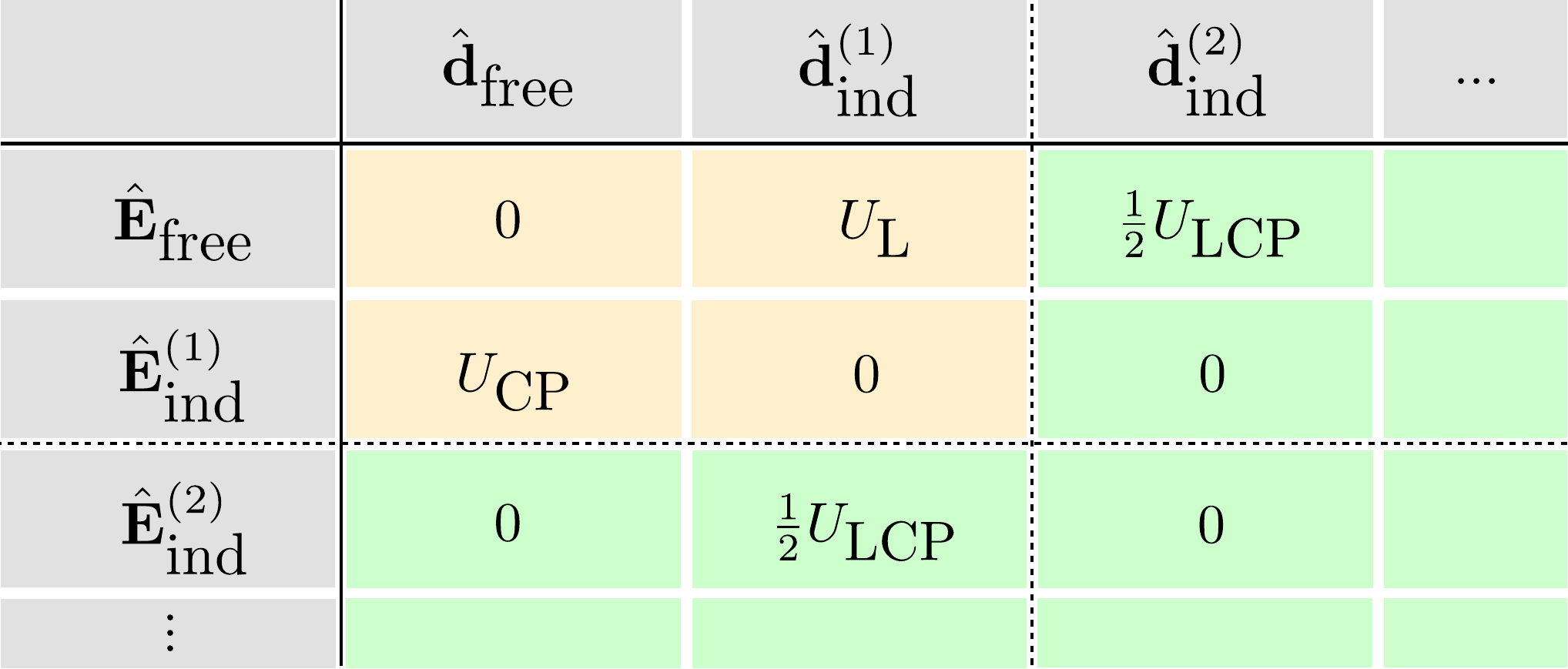}}
\caption{Summary of contributions to the total potential $U$. The contributions to lowest order (yellow) form the ordinary CP potential $U_{\textrm{CP}}$ and the laser-light potential $U_{\textrm{L}}$. Higher-order terms (green) build up the non-additive laser-induced CP potential $U_{\textrm{LCP}}$.}
\label{fig:Cascade}
\end{table}
The potential $U = - \frac{1}{2} \langle \hat{\vec{d}} \left( t \right) \cdot \hat{\vec{E}} \left( \vec{r}_{\textrm{A}}, t \right) \rangle$ at the atom's position in lowest order comprises four terms. The first term $\langle \hat{\vec{d}}_{\textrm{free}} \left( t \right) \cdot \hat{\vec{E}}_{\textrm{free}} \left( \vec{r}_{\textrm{A}}, t \right) \rangle$ contains the free dipole moment and the free electric field. For $\vec{r}' \in V_{\textrm{S}}$, this expression leads to the vanishing expectation value of the free dipole moment $\langle \hat{\vec{d}}_{\textrm{free}} \left( t \right) \rangle = 0$. In case of $\vec{r}' \notin V_{\textrm{S}}$, this term vanishes as well according to Eqs.~\eqref{eq:Quantum State} and \eqref{eq:Electric Field Quantum State}. The standard, undriven CP potential is obtained from the term $\langle \hat{\vec{d}}_{\textrm{free}} \left( t \right) \cdot \hat{\vec{E}}_{\textrm{ind}} \left( \vec{r}_{\textrm{A}}, t \right) \rangle$ \cite{Buhmann:2004}
\begin{multline}
U_{\textrm{CP}} \left( \vec{r}_{\textrm{A}} \right) = \frac{\hbar \mu_0}{2 \pi} \int \limits^{\infty}_0{\dif \xi \xi^2 \alpha_n \left( \mi \xi \right) \textrm{Tr} \tens{G}^{(\textrm{S})} \left( \vec{r}_{\textrm{A}}, \vec{r}_{\textrm{A}}, \mi \xi \right)}\\
- \frac{\mu_0}{3} \sum\limits_{k<n} {\tilde{\omega}^2_{nk} \left| \vec{d}_{nk} \right|^2 \textrm{Tr} \left[ \textrm{Re} \tens{G}^{(\textrm{S})} \left( \vec{r}_{\textrm{A}}, \vec{r}_{\textrm{A}}, \tilde{\omega}_{nk} \right) \right]}.
\label{eq:Casmir-Polder Potential}
\end{multline}
$\tens{G}^{(\textrm{S})} \left( \vec{r}_{\textrm{A}}, \vec{r}_{\textrm{A}}, \tilde{\omega}_{nk} \right)$ is the scattering part of the classical Green's tensor $\tens{G} \left( \vec{r}_{\textrm{A}}, \vec{r}_{\textrm{A}}, \tilde{\omega}_{nk} \right)$. The CP potential can be split into resonant and off-resonant contributions, where the ground-state only shows the latter. The third term of the total potential $U$ results in the laser-light potential $\langle \hat{\vec{d}}_{\textrm{ind}} \left( t \right) \cdot \hat{\vec{E}}_{\textrm{free}} \left( \vec{r}_{\textrm{A}}, t \right) \rangle = \langle \hat{\vec{d}}_{\textrm{ind}} \left( t \right) \rangle \cdot \vec{E} \left( \vec{r}_{\textrm{A}}, t \right)$ from the coherent time-averaged electric field $\left( \vec{r}' \in V_{\textrm{S}} \right)$
\begin{equation}
U_{\textrm{L}} \left( \vec{r}_{\textrm{A}} \right) = -\frac{1}{4} \alpha_n \left( \omega_{\textrm{L}} \right) \vec{E}^2 \left( \vec{r}_{\textrm{A}} \right)
\label{eq:Laser Light Potential}
\end{equation}
and the fourth term of the total potential $\langle \hat{\vec{d}}_{\textrm{ind}} \left( t \right) \cdot \hat{\vec{E}}_{\textrm{ind}} \left( \vec{r}_{\textrm{A}}, t \right) \rangle$ vanishes again both for $\vec{r}' \in V_{\textrm{S}}$ and $\vec{r}' \notin V_{\textrm{S}}$.

We are interested in a higher-order iteration, where the induced part of the dipole moment itself depends on the induced part of the electric field, while the induced electric field itself contains the induced dipole moment. Combining the two high-order perturbative expressions for $\hat{\vec{d}}$ and $\hat{\vec{E}}$ leads to the final result for the driven CP potential
\begin{align}
&U_{\textrm{LCP}} \left( \vec{r}_{\textrm{A}} \right)\notag \\
&= -\frac{1}{2} \langle \hat{\vec{d}}_{\textrm{ind}} \left( t \right) \sprod \hat{\vec{E}}^{(2)}_{\textrm{ind}} \left( \vec{r}_{\textrm{A}}, t \right) \rangle - \frac{1}{2} \langle \hat{\vec{d}}^{(2)}_{\textrm{ind}} \left( t \right) \sprod \hat{\vec{E}}_{\textrm{free}} \left( \vec{r}_{\textrm{A}}, t \right) \rangle\notag \\
&= - \frac{\mu_0 \omega^2_{\textrm{L}}}{2} \alpha^2_n \left( \omega_{\textrm{L}} \right) \vec{E} \left( \vec{r}_{\textrm{A}} \right)  \sprod \mathrm{Re} \tens{G}^{(\textrm{S})} \left( \vec{r}_{\textrm{A}}, \vec{r}_{\textrm{A}}, \omega_{\textrm{L}} \right) \sprod \vec{E} \left( \vec{r}_{\textrm{A}} \right).
\label{eq:Driven Casimir-Polder Potential Perturbation}
\end{align}
This expression contains induced dipole moments and induced electric fields, both of second order. The CP force corresponding to the potential \eqref{eq:Driven Casimir-Polder Potential Perturbation} is computed by taking the gradient of the potential $\vec{F}_{\textrm{LCP}} \left( \vec{r}_{\textrm{A}} \right) = -  {\nabla}_{\textrm{A}} U_{\textrm{LCP}} \left( \vec{r}_{\textrm{A}} \right)$ and can be expressed using the two contributions $ {\nabla} \langle \hat{\vec{d}}^{(2)}_{\textrm{ind}} \sprod \hat{\vec{E}}_{\textrm{free}} \left( \vec{r} \right) \rangle_{\vec{r} = \vec{r}_{\textrm{A}}} +  {\nabla} \langle \hat{\vec{d}}_{\textrm{ind}} \sprod \hat{\vec{E}}^{(2)}_{\textrm{ind}} \left( \vec{r} \right) \rangle_{\vec{r} = \vec{r}_{\textrm{A}}}$, where one can use the relation $\left.  {\nabla} \vec{E} \left( \vec{r}_{\textrm{A}} \right) \sprod \vec{E} \left( \vec{r} \right) \right|_{\vec{r} = \vec{r}_{\textrm{A}}} = \frac{1}{2}  {\nabla}_{\textrm{A}} \vec{E}^2 \left( \vec{r}_{\textrm{A}} \right)$ and the symmetry of the Green's tensor ${\nabla} \tens{G}^{(\textrm{S})}\left( \vec{r}, \vec{r}_{\textrm{A}} \right) |_{\vec{r} = \vec{r}_{\textrm{A}}}= \frac{1}{2}  {\nabla}_{\textrm{A}} \tens{G}^{(\textrm{S})} \left( \vec{r}_{\textrm{A}}, \vec{r}_{\textrm{A}} \right)$.

Figure \ref{fig:PlotTwo} shows the difference between the laser-induced CP potential and the standard CP potential $U_{\textrm{CP}}$ for a perfectly conducting mirror, whose the Green's tensor is well-known.
The laser-induced potential for a constant laser potential is similar to the CP potential of an excited atom. The strongest addition effect of $U_{\textrm{LCP}}$ is seen in the nonretarded regime, which we investigate further for a more realistic evanescent laser field.

For simplicity we first consider a two-level atom a distance $z$ away from a dielectric half-space described by $\textrm{s}$ and $\textrm{p}$ polarised reflection coefficients $r_{\textrm{s}}$ and $r_{\textrm{p}}$. We work in the non-retarded (small $z$) limit and assume isotropic polarizability subject to $\tilde{\omega}_{10} \gg \Delta$, in which case Eq.~\eqref{eq:Atomic Polarizability} reads
\begin{equation}
\alpha_n \left( \omega_{\textrm{L}} \right) = \frac{\alpha_{\textrm{DC}} \tilde{\omega}_{10}}{2 \left( \tilde{\omega}_{10} - \omega_{\textrm{L}} \right)} \tens{1} = -\frac{\alpha_{\textrm{DC}} \tilde{\omega}_{10}}{2 \Delta} \tens{1},
\end{equation}
where we have used the expression $\alpha_{\textrm{DC}} = 2 \left| \vec{d}_{nk} \right|^2 / \left( 3 \hbar \tilde{\omega}_{10} \right)$ and defined a detuning $\Delta = \omega_{\textrm{L}} - \tilde{\omega}_{10}$. By expressing the electric field $\left| \vec{E} \left( \vec{r}_{\textrm{A}} \right) \right|$ in terms of the intensity $I = \frac{1}{2} \epsilon_0 c \left| \vec{E} \left( \vec{r}_{\textrm{A}}, t \right) \right|^2$, the laser-induced CP potential $U_{\textrm{LCP}}$ \eqref{eq:Driven Casimir-Polder Potential Perturbation} in this limit can be represented as a function of the laser intensity $I$ and the detuning $\Delta$ between the atomic transition frequency $\tilde{\omega}_{10}$ and the laser frequency $\omega_{\textrm{L}}$. In the nonretarded limit $U_{\textrm{LCP}}$ reads
\begin{equation}
U_{\textrm{LCP}} \left( z_{\textrm{A}} \right) = - \frac{\tilde{\omega}^2_{10} \alpha^2_{\textrm{DC}} I \left( z_{\textrm{A}} \right)}{128 \epsilon^2_0 \pi c \Delta^2 z^3_{\textrm{A}}} \textrm{Re} \left( r_{\textrm{p}} \right),
\label{eq:U_LCP}
\end{equation}
where $r_{\textrm{p}} = \left( \epsilon - 1 \right) / \left( \epsilon + 1 \right)$ is the nonretarded limit of the p-polarized reflection coefficient with permittivity $\epsilon$. We relate the real part of $r_{\textrm{p}}$ to the quality factor $Q$ for our system: $\textrm{Re} \left( r_{\textrm{p}} \right) = \pm Q$. By taking the Drude--Lorentz model as a basis, the quality factor reads $Q \approx \omega_{\textrm{S}}/2 \gamma$ with the plasmon resonance frequency $\omega_{\textrm{S}} = \sqrt{\omega^2_0 + \frac{1}{2} \omega^2_{\textrm{P}}}$ containing the resonance frequency $\omega_0$, the plasma frequency $\omega_{\textrm{P}}$ and the damping constant $\gamma$, cf. Ref.~\cite{Chang:2013}. Then, our result \eqref{eq:U_LCP} can also be expressed in the following suggestive way
\begin{equation}
U_{\textrm{LCP}} \hbar \Delta = U_{\textrm{L}} U_{\textrm{CP}} Q
\end{equation}
so that in this regime the non-additive CP potential is seen to depend on the product of the additive potentials, modulated by $Q$ and the detuning. As shown in Refs.~\cite{Esslinger:1993, Bartolo:2016, Slama:2014, Dung:2001}, surface plasmon resonances, connected in the context of CP potentials, can be used to produce a maximum Purcell enhancement factor, or quality factor $Q$, of up to $60$.

In order to produce concrete predictions, we consider the setup of Refs.~\cite{Bender:2010} and \cite{Bender:2014}, where an evanescent wave is created close to a surface creating a repulsive dipole potential
\begin{equation}
U_{\textrm{L}} \left( z_{\textrm{A}} \right) = C_0 P \exp\left( -2 z_{\textrm{A}} /z_0 \right)
\end{equation}
with a decay parameter $z_0 = 430$ nm and laser power $P$. The factor $C_0$ is computed as in Refs.~\cite{Bender:2014, Bender:2010} using the transition frequency of Rb $\tilde{\omega}_{10} = 2.37 \times 10^{15}$ rad/s and the dipole moment $\left| \vec{d}_{nk} \right| = 2.53 \times 10^{-29}$ Cm. Our assumption of a single transition frequency is justified by the small detuning of $\Delta = \omega_{\textrm{L}} - \tilde{\omega}_{10} = 2 \pi \times 10^8$ rad/s. Using an optical diffraction coefficient of $n=1.512$ we obtain a value of $C_0 = 4.51 \times 10^{-23}$ J/W. Similar to the values in Ref.~\cite{Bender:2010} we assumed the laser to have an elliptical beam waist with $w_x = 170 \: \mu$m and $w_y = 227 \: \mu$m and a laser power of $P=39 \; \mu$W. The resulting potential $U_{\textrm{L}}$ is to be compared with the nonretarded CP potential for ground-state atoms $U_{\textrm{CP}} \left( z_{\textrm{A}} \right) = -C_3/z^3_{\textrm{A}}$ with
\begin{equation}
C_3 = \frac{\hbar}{16 \pi^2 \epsilon_0} \int\limits^{\infty}_0 \dif \omega {\alpha \left( \mi \omega \right) r_{\textrm{p}} \left( \mi \omega \right)} = \frac{\alpha_{\textrm{DC}} \hbar \tilde{\omega}_{10}}{32 \pi \epsilon_0}
\label{eq:C_3}
\end{equation}
from Ref.~\cite{Perreault:2008}, which only give an estimate for the nonretarded limit but can be generalized using Eq.~\eqref{eq:Driven Casimir-Polder Potential Perturbation}. The final equality in Eq.~\eqref{eq:C_3} holds for a perfect conductor and an undamped atom modeled by the Lorentz model.

\begin{figure}
\centerline{\includegraphics[width=\columnwidth]{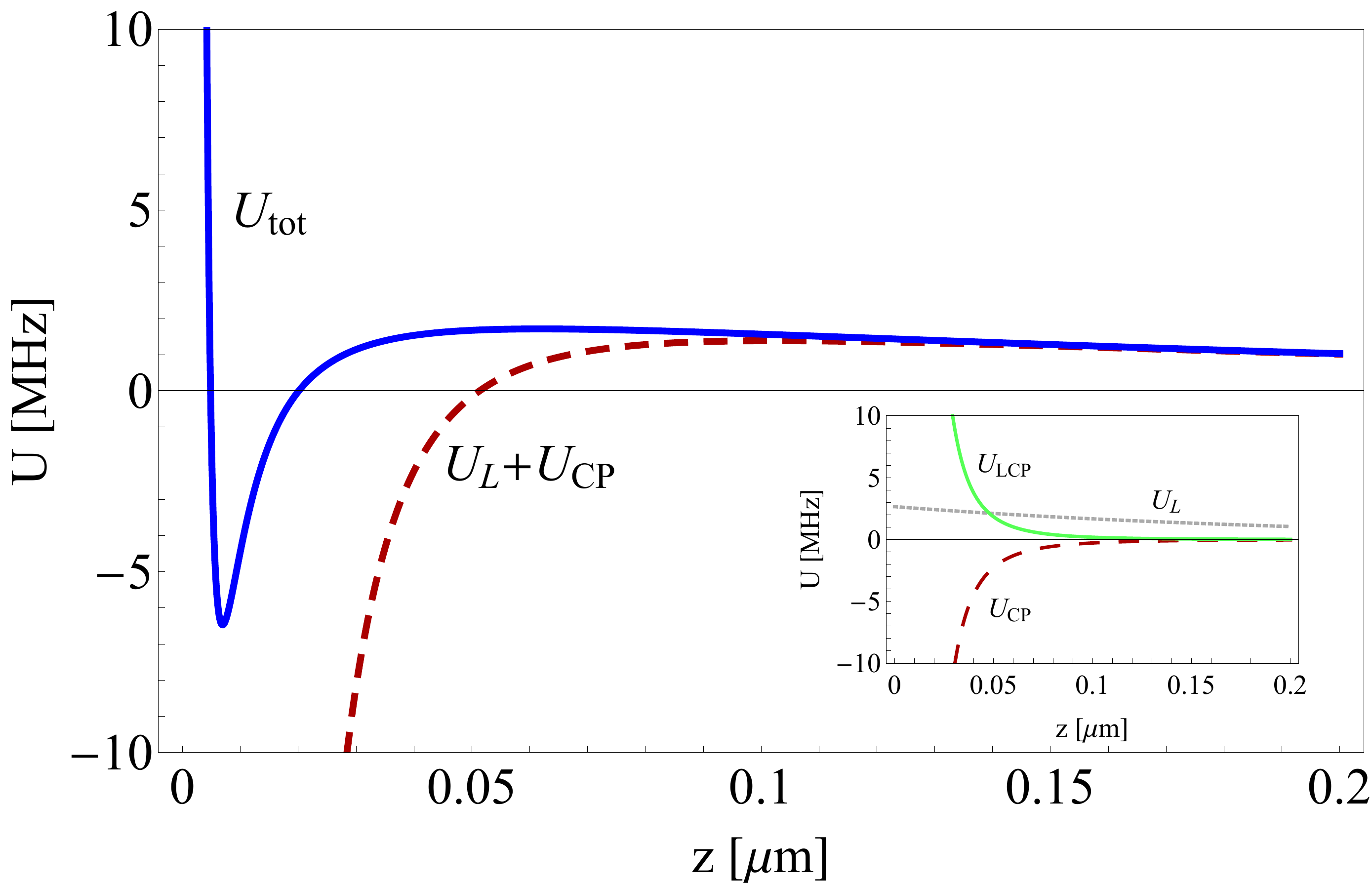}}
\caption{Comparison between the CP potential $U_{\textrm{CP}}$, the evanescent laser potential $U_{\textrm{L}}$, the nonadditive potential $U_{\textrm{LCP}}$ and the total potential $U_{\textrm{tot}}$ at a laser power of $P = 39 \: \mu$W with an elliptical beam waist with $w_x = 170 \: \mu$m and $w_y = 227 \: \mu$m and a plasmonic enhancement factor of $60$.}
\label{fig:PlotU}
\end{figure}

Figure~\ref{fig:PlotU} shows the CP potential $U_{\textrm{CP}}$, the evanescent laser potential $U_{\textrm{L}}$, the nonadditive potential $U_{\textrm{LCP}}$ and the sum of all of these terms $U_{\textrm{tot}}$. We used a factor for Q of $60$ and a laser power value of $P = 39 \; \mu$W. If one adds the nonadditive potential $U_{\textrm{LCP}}$ to the traditionally studied contributions $U_{\textrm{L}}$ and $U_{\textrm{CP}}$, one observes the emergence of a pronounced dip which could serve as a trapping potential for the atoms.

\begin{figure}
\centering
\begin{overpic}[width=\columnwidth]{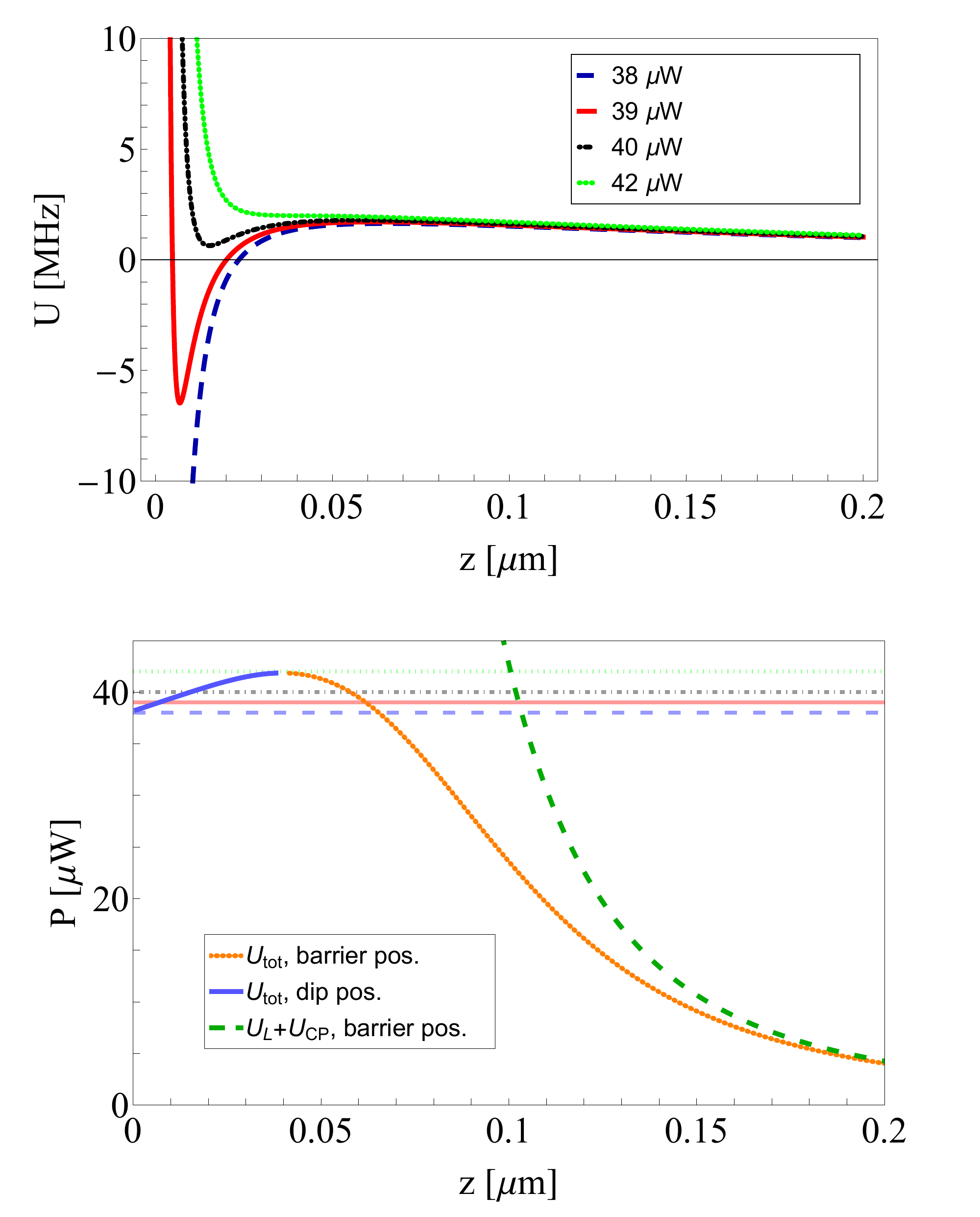}
\put(0,100){a)}
\put(0,50){b)}
\end{overpic}
\caption{Nonadditive potential $U_{\textrm{LCP}}$ for four different values of the laser power $P$. A change from an attractive to a repulsive potential can be observed at a certain laser power $P$ with an elliptical beam waist with $w_x = 170 \: \mu$m and $w_y = 227 \: \mu$m. The second part shows the existence and position of minima and maxima as a function of laser power, illustrating the transition from a trap-like potential to a barrier-like one as a function of the distance of the atom and the surface.}
\label{fig:PlotL}
\end{figure}

Figure~\ref{fig:PlotL} a) compares the nonadditive potential $U_{\textrm{LCP}}$ for several values of the laser power $P$, where one observes a change from attractive to repulsive values at a certain laser power as shown in Fig.~\ref{fig:PlotL} b). There it is seen that according to previous theory a barrier forms and moves closer to the surface as the power is increased. By contrast, once our correction is included the position of this barrier changes, then becomes a dip for sufficient laser power.

In this paper we have derived and theoretically evaluated a non-additive laser-induced CP potential. In a perturbative approach the electric field and the dipole moment were each split into a free contribution and an induced contribution, each of which depends on the other. In this way the laser light potential and the standard CP potential are reproduced as lowest-order terms. The higher-order correction term, where the induced components depend on other induced components, leads to the nonadditive potential. We have shown that this term makes a significant contribution under certain experimental conditions. If the laser power is high enough and in combination with an additional enhancement by a surface plasmon resonance, the occurrence and position of barriers and minima in the total potential can significantly change, leading to local minima which can be used to trap atoms near surfaces. 

\begin{acknowledgments}We acknowledge helpful discussions with Diego Dalvit, Francesco Intravaia and Ian Walmsley. This work was supported by the German Research Foundation (DFG, Grants BU 1803/3-1 and GRK 2079/1). S.Y.B is grateful for support by the Freiburg Institute of Advanced Studies.\end{acknowledgments}


%

\end{document}